\documentclass[12pt]{article}
\begin{document}

%%%%%%%%%%%%%%%%%%%%%%%%%%%%%%%%%%%%%%%%%%%%%%%%

\def\a{\alpha}
\def\b{\beta}
\def\g{\gamma}
\def\G{\Gamma}
\def\d{\delta}
\def\D{\Delta}
\def\e{\epsilon}
\def\h{\hbar}
\def\ve{\varepsilon}
\def\z{\zeta}
\def\t{\theta}
\def\vt{\vartheta}
\def\r{\rho}
\def\vr{\varrho}
\def\k{\kappa}
\def\l{\lambda}
\def\L{\Lambda}
\def\m{\mu}
\def\n{\nu}
\def\o{\omega}
\def\O{\Omega}
\def\s{\sigma}
\def\vs{\varsigma}
\def\S{\Sigma}
\def\vphi{\varphi}
\def\av#1{\langle#1\rangle}
\def\pa{\partial}
\def\na{\nabla}
\def\hg{\hat g}
\def\un{\underline}
\def\ov{\overline}
\def\cF{{\cal F}}
\def\cG{{\cal G}}
\def\cN{{\cal N}}
\def\Hsl{H \hskip-8pt /}
\def\Fsl{F \hskip-6pt /}
\def\cFsl{\cF \hskip-5pt /}
\def\ksl{k \hskip-6pt /}
\def\pasl{\pa \hskip-6pt /}
\def\tr{{\rm tr}}
\def\tcF{{\tilde{{\cal F}_2}}}
\def\tg{{\tilde g}}
\def\shalf{\frac{1}{2}}
\def\nn{\nonumber \\}
\def\w{\wedge}
\def\ra{\rightarrow}
\def\la{\leftarrow}
\def\be{\begin{equation}}
\def\ee{\end{equation}}
\def\ad{\bigtriangledown}
\newcommand{\brr}{\begin{eqnarray}}
\newcommand{\err}{\end{eqnarray}}

%%%%%%%%%%%%%%%%%%%%%%%%%%%%%%%%%%%%%%%%%%%%%%%%%%%%%%%%%%%%%%%%%%%%%%

\begin{titlepage}

%%%%%%%%%%%%%%%%%%%%%%%%%%%%%%%%%%%%%%%%%%%%%%%%%%%%%%%%%%%%%%%%%%%%%%

\setcounter{page}{0}

\begin{flushright}
COLO-HEP-42 \\
hep-th/0007nnn \\
July 2000
\end{flushright}

\vspace{5 mm}
\begin{center}
{\large {\bf The Cosmological Constant in Calabi-Yau 3-fold Compactifications
of the Horava-Witten Theory}}
\vspace{20 mm}

{\large S. P. de Alwis and N. Irges\footnote{e-mail:
dealwis@pizero.colorado.edu,
irges@pizero.colorado.edu}}\\
{\em Department of Physics, Box 390,
University of Colorado, Boulder, CO 80309.}\\
\vspace{5 mm}
\end{center}

\vspace{10 mm}

\centerline{{\bf{Abstract}}}
Brane world scenarios offer a way of setting 
the cosmological constant to zero after supersymmetry breaking 
provided there is a sufficient number of adjustable 
integration constants/parameters.
In the case of the Horava-Witten theory compactified on a Calabi-Yau
threefold, we argue that it is difficult to find enough freedom to
get a zero (or small) cosmological constant after supersymmetry
breaking.

%%%%%%%%%%%%%%%%%%%%%%%%%%%%%%%%%%%%%%%%%%%%%%%%%%%%%%%%%%%%%%%%%%%%%%%
\end{titlepage}
%%%%%%%%%%%%%%%%%%%%%%%%%%%%%%%%%%%%%%%%%%%%%%%%%%%%%%%%%%%%%%%%%%%%%%%%

\newpage
\renewcommand{\thefootnote}{\arabic{footnote}}
\setcounter{footnote}{0}

%%%%%%%%%%%%%%%%%%%%%%%%%%%%%%%%%%%%%%%%%%%%%%%%
\section{Introduction}
%%%%%%%%%%%%%%%%%%%%%%%%%%%%%%%%%%%%%%%%%%%%%%%%

One of the phenomenologically most relevant corners of the string/M-theory
moduli space is the Horava-Witten theory \cite{HW}. It corresponds to
compactifying M-theory on $R^{10}\times S^1/Z_2$, which results in an eleven
dimensional ``brane world'', with two ten dimensional branes sitting at
the fixed points of the orbifold, each of them supporting an $E_8$ gauge
theory. Compactifying six of the dimensions of $R^{10}$ on  
a Calabi-Yau three-fold (${\rm CY}_3$), results in a five dimensional brane world with 
an ${\cal N}=2$ gauged supergravity in the bulk and an 
${\cal N}=1$ supersymmetric $E_8$ gauge theory on each three-brane at
the fixed points. A detailed derivation of the five 
dimensional low energy effective
action for these compactifications was given in \cite{M}.  

Such five dimensional brane world compactifications, 
besides their obvious phenomenological
relevance due to the four dimensional ${\cal N}=1$ gauge theories
coupled to chiral matter,   
offer a means for setting the four dimensional cosmological 
constant (CC) \cite{cc} \cite{DeW} to zero, provided there 
are adjustable integration constants/parameters. Indeed, in 
another class of brane world scenarios, 
namely compactifications of type II string
theory on curved spaces (such as $S^5$ or deformations of it) with 
the gauge theory coming from D-branes sitting at 
orientifold fixed planes, it is indeed possible to 
argue that one can get flat space
after supersymmetry breaking by adjusting 
integration constants \cite{sda}. In this paper, we
discuss the same issue in the context of the Horava-Witten theory.

It was shown in the third paper of reference 
\cite{HW} that a consistent compactification
of M-theory ala Horava-Witten on a Calabi-Yau three-fold 
requires a non-trivial non-zero mode flux 
for the four form field strength ${\cal G}$. This corresponds to wrapped
M5-branes around non-trivial $(1,1)$ cycles of the Calabi-Yau and one of
the consequences of this background flux is a non-zero potential for the
volume and shape moduli \cite{M}. 

One could also consider the existence, for example
of instanton corrections coming from  (Euclidean) M5-branes wrapped around the
Calabi-Yau manifold. Indeed,
this seems to be the case, as discussed in \cite{inst2}, 
even though the exact form of this correction to the potential is not
known. What we do know, however, is that the volume modulus dependence
is the same as for the potential from the compactification, a fact that
can be easily justfied once we realize that both contributions can be
seen as arising from gauging Abelian isometries of the universal
hypermultiplet coset space \cite{inst1}, \cite{inst2}. The total potential 
due to M5-branes can be therefore written as 
\be V_{M5}= e^{2\vphi} (V_{(1,1)}+V_{(3,3)}),\label{VM5}\ee 
where $\vphi$ is the volume modulus, 
$V_{(1,1)}$ and $V_{(3,3)}$ correspond to
M5-branes wrapped around $(1,1)$ and the $(3,3)$ cycle respectively.

In section 2, we  rederive $V_{(1,1)}$ 
following \cite{M} ignoring for simplicity instanton contributions. In section 3  we discuss the cosmological
constant problem from the point of view of the 
five dimensional brane world after supersymmetry breaking.  We will conclude that 
after supersymmetry breaking there do not exist flat 
(or nearly flat) brane solutions to
these equations. In section 4, we summarize our conclusions.

%%%%%%%%%%%%%%%%%%%%%%%%%%%%%%%%%%%%%%%%%%%%%%%%
\section{Horava-Witten on ${\rm CY}_3$ }
%%%%%%%%%%%%%%%%%%%%%%%%%%%%%%%%%%%%%%%%%%%%%%%%

The bosonic part of the low energy effective
action of $M$-theory is the bosonic part of the eleven dimensional
supergravity of \cite{11sugra}:
\be \label{11action}S=-{1\over {2{\k^{(11)}}^2}}
\Bigl\{\int d^{11}x\sqrt{-g^{(11)}}
{\cal R}^{(11)}-{1\over 2}\int \Bigl(
{\cal G}\wedge *{\cal G}+{1\over 3}{\cal C}\wedge
{\cal G}\wedge {\cal G}\Bigr)\Bigr\},\ee
with ${\cal C}$ the 3-form of 11 dimensional supergravity and ${\cal
G}=d{\cal C}$. This action, when supplemented by appropriate boudary
terms, describes the (low energy) strong coupling limit of the heterotic string
theory, of Horava and Witten.

We first consider the effective action obtained by reducing (\ref{11action}) on
a Calabi-Yau 3-fold \cite{11CY}. The appropriate metric ansatz is
\be {ds^{(11)}}^2={ds^{(5)}_{str}}^2+{ds^{(6)}_{CY}}^2.\ee
The subscript $''str''$ indicates string frame and
superscripts indicate the dimensionality.
We assume, following \cite{HW} \cite{M},
that the ``standard embedding'' requires in the Horava-Witten picture a
non-trivial ${\cal G}$-flux of the form
\be {\cal G}={i\over {2{\cal V}}}n_iG^{ij} (*\omega_j),\ee
where the $n_i$ are integers.
The form of the ${\cal G}$-flux for non-standard embeddings and additional five
branes in the bulk is similar; the only modification is that the
$n_i$ in the above is replaced by a sum over fluxes \cite{nonstandard}.
The low energy spectrum includes the
five dimensional graviton multiplet, $h^{1,1}-1$ vector multiplets
containing one shape modulus scalar ($b^i$) each ($i=1... h^{1,1}$
but there is one constraint, see below), 
and the universal hypermultiplet
that includes the volume modulus ($\vphi$), the scalar dual to the 
five dimensional three-form
($\s$)  and a pair of complex scalars ($\xi$ and ${\bar \xi}$).
The metric on the vector moduli space is
\be \label{metric} G_{ij}(a)={i\over {2{\cal V}}}\int_{CY}
\omega_i\wedge *\omega_j=-{1\over 2}\pa_i\pa_j\ln {\cal V}(a),\ee
The K${\rm {\ddot a}}$hler form is $J=a^i\omega_i$, with
the $\omega_i$ a basis of $h^{1,1}$ 2-forms and $a^i$ 
are the $h^{1,1}$ K${\rm {\ddot a}}$hler moduli of the Calabi-Yau.
The Calabi-Yau volume
${\cal V}$ is
\be \label{CYvolume} {\cal V}(a)={1\over 3!}\int_{CY}J\wedge J \wedge J={1\over
6}c_{ijk}a^ia^ja^k,\ee
with $c_{ijk}$ the intersection numbers of the Calabi-Yau.

For our purposes, it is sufficient to consider the part of the effective action
which is 5D gravity coupled to the scalars of the vector multiplets 
and the volume modulus (breathing mode) of the universal hypermultiplet.
The corresponding five dimensional string frame bulk effective action resulting
from the first two terms of
(\ref{11action}), is
\be \label{5actionstr}S = -{1\over {2{\k^{(5)}}^2}}\int d^5x\sqrt{-g_{str}^{(5)}}
\Bigl[{\cal V} {\cal R}^{(5)}+\bigl({\cal V}
G_{ij}(a)+\pa_i\pa_j{\cal V} \bigr)
\pa_{I}a^i\pa^{I}a^j-{1\over
{4{\cal V} }}G^{ij}(a)n_in_j\Bigr].\ee
The five dimensional index is $I=\{\mu,r\}$. We have neglected the terms coming
from the Chern-Simons term in (\ref{11action}) since they are irrelevant
to our dicussion. After the Weyl rescaling 
${ds^{(5)}_E}^2={\cal V}^{2\over 3}{ds^{(5)}_{str}}^2$ and separation of
the volume modulus from the shape moduli 
which can be done by defining $b^i=a^i{\cal
V}^{-1/3}$, we arrive at a bulk action of the form  
\be S_{bulk} = -{1\over {2{\k^{(5)}}^2}}
\int d^5x \sqrt{-g_{E}^{(5)}}\Bigl[{\cal R}^{(5)}
-G_{ij}(b)\pa_{I}b^i\pa^{I}b^j-
{1\over 2}\pa_{I}\vphi\pa^{I}\vphi -{1\over
4}e^{2\vphi}G^{ij}(b)n_in_j+\lambda(c_{ijk}b^ib^jb^k-6)\Bigr],\ee
where we have defined ${\cal V}=e^{-\vphi}$ and 
$\lambda$ is a Lagrange multiplier.
This is not the whole relevant Horava-Witten 5D action, 
because we have not taken into account yet the
Horava-Witten Wall/M5-brane action sitting at the fixed points of the orbifold.
The additional brane terms are 
\be S_{branes}={1\over {2\k^{(5)}}^2}\int 
d^4x \sqrt{-g_E^{(4)}}T_{1}(\vphi )\delta(r)+
{1\over {2\k^{(5)}}^2}\int d^4x \sqrt{-g_E^{(4)}}T_{2}(\vphi )\delta(r-\pi R),\ee
with $g_E^{(4)}$ the induced metric on the brane (we will assume static
gauge and ignore fluctuations) and $T_{1,2}$ are the tensions of the
branes. As we mentioned, these 3-branes arise from the
M5-branes of the original 11 dimensional theory that are wrapped
around non-trivial 2-cycles of the Calabi-Yau 
and the brane tensions at the string scale, i.e. before
supersymmetry breaking, are simply 
proportional to the volume of the 2-cycle \cite{M}, \cite{st}:
\be \label{tension}T_1=-T_{2}(\vphi)={2\over 1!}\int J=
2e^{\vphi}\z.\ee
Defining $n^2\equiv G^{ij}n_in_j=n_i n^i$ 
and eliminating the Lagrange multiplier, 
we can write the equations of motion for the $b^i$ as 
\begin{eqnarray} & -{1\over 4}e^{2\vphi}(\pa_k n^2)+
{1\over 6}e^{2\vphi}b_kb^l(\pa_l n^2) & \nonumber \\
& -{3\over 2}c_{ijk}(\ad_{\a}b^i)(\ad^{\a}b^j)
+{2\over 3}c_{ijl}b^lb_k(\ad_{\a}b^i)(\ad^{\a}b^j)+
{4\over 3}b_kb_i(\ad_{\a}\ad^{\a}b^i) & \nonumber \\
& = \label{eqX} e^{\vphi}(\d(r)-\d(r-\pi R))\Bigl({4\over
3}n_lb^lb_k-2n_k\Bigr). &
\end{eqnarray}
To simplify these equations, we will assume that the $b^i$ take constant
values in the vacuum. All the terms with the covariant derivatives
drop out, the equations of motion for $b^i$ decouple from the
equation of motion for $\vphi$ and then 
the unique ansatz that solves the remaining of 
(\ref{eqX}) is \cite{M}, \cite{st}
\be \z b_k={3\over 2}n_k,\label{Taylor}\ee
where $\zeta\equiv n_lb^l$.
By solving the above system of equations, we obtain the values
that the squashing modes take in the vacuum.
Notice that for the ansatz (\ref{Taylor}), the term multiplying the
$\d$-functions in (\ref{eqX}) vanishes, which is a necessary condition
for a consistent solution with constant $b^i$.
Now we turn to the breathing mode $\vphi$.

The bulk action, in the Einstein frame,
with all the moduli besides the breathing mode $\vphi$ stabilized, is
\be\label{roundaction} S_{bulk} = -{1\over {2{\k^{(5)}}^2}}
\int d^5x\sqrt{-g_E^{(5)}}\Bigl[{\cal R}^{(5)}-\shalf
(\pa\vphi )^2-{1\over 6}\z^2e^{2\vphi}\Bigr].\ee
In the above, the constant $\z^2$ in front of the potential is determined
in terms of the integer $n_i$ 
(therefore it is not a continuous quantity) and it corresponds
to contributions from fluxes associated with wrapping M5 branes
around the non trivial $(1,1)$ cycles of the Calabi Yau.
>From the 5D gauged supergravity point of view, 
the potential comes from gauging 
the $U(1)$ isometry of the universal hypermultiplet moduli space 
$SU(2,1)/U(2)$ corresponding to a shift symmetry of the scalar $\s$. 
It is interesting to note here that the universal hypermultiplet moduli
space has more $U(1)$ isometries that could be gauged \cite{inst1},
associated with shift symmetries of $\xi$ and ${\bar \xi}$.
Gauging all of the $U(1)$ isometries, 
in principle, can produce additional terms in the potential.
>From the M-theory point of view, the additional pieces correspond to
M5-branes wrapped around the $(3,3)$ cycle (i.e. the whole Calabi-Yau) 
and/or to M2-branes wrapped around $(3,0)$ 
and $(0,3)$ cycles of the Calabi-Yau \cite{inst2}. 
The potential in (\ref{roundaction}) is therefore 
the potential for ${\rm CY}_3$ compactifications 
of the Horava-Witten theory for constant shape moduli and 
without an M5/M2 instanton gas. In the following, we will neglect
instanton contributions, since our subsequent arguments about the
cosmological constant are not affected by their presence. 
%%%%%%%%%%%%%%%%%%%%%%%%%%%%%%%%%%%%%%%%%%%%%%%%
\section{Supersymmetry Breaking and the Cosmological Constant }
%%%%%%%%%%%%%%%%%%%%%%%%%%%%%%%%%%%%%%%%%%%%%%%%
The equations of motion from action (\ref{roundaction}), 
for $\vphi=\vphi(r)$ and for
\be
{ds_E^{(5)}}^2=e^{2A(r)}\eta_{\mu\nu}
dx^{\mu}dx^{\nu}+{(dr)}^2,\label{eq:metric}\ee
are
\be \vphi ''+4A'\vphi '={{{\partial
V(\vphi)}\over {\partial {\vphi}}}}\label{eq:Beq1}\ee
\be A''=-{1\over 6}\vphi'^2\label{eq:Beq2}\ee
\be A'^2=-{1\over 12}V(\vphi)+{1\over 24}\vphi'^2, \label{eq:Beq3}\ee
where $V(\vphi)\equiv e^{2\vphi}V_{(1,1)}=e^{2\vphi}{1\over 6}\z^2$.
The prime denotes differentiation with respect to $r$.
The first order set of equations corresponding to the above 
second order set, in terms of the superpotential $W$, is
\be \vphi'={{{{\partial W}\over
{\partial {\vphi}}}}}\label{eq1}\ee
\be A'=-{1\over 6}W\label{eq2}\ee
\be V(\vphi)={1\over 2}\Bigl({{{\partial W}\over {\partial
{\vphi}}}}\Bigr)^2-{1\over 3}W^2.\label{VW}\ee
We note here that the proper ansatz for the five dimensional
metric would be to replace the flat Minkowski metric
$\eta_{\mu\nu}$ by $g^{(A)dS}_{\mu\nu}$ (i.e. a metric for 
deSitter or Anti-deSitter four-space) but since the (possibly) measured value of
the CC is many tens of orders of magnitude smaller 
than all relevant scales in this
analysis, this is not a meaningful difference.
We can rewrite (\ref{VW}) as
\be \Bigl({{\dot W}\over {\sqrt{2V}}}\Bigr)^2-
\Bigl({{W}\over {\sqrt{3V}}}\Bigr)^2=1,\ee
where the dot stands for differentiation
with respect to $\vphi$. The ansatz
\be {\dot W}={\sqrt{2V}}\cosh{f(\vphi)}\;\;\;\; {\rm and}\;\;\;\;
W={\sqrt{3V}}\sinh{f(\vphi)}\label{elliptic}\ee
allows us to separate variables, so that
combining the above, we can integrate over $\vphi$ and $f$:
\be \label{integ}\int {d\vphi}=\int {df\over {\sqrt{2\over 3}-\tanh {f}}}. \ee
The integral yields
\be \label{sol1} c_1e^{-\vphi}=(\tanh{f}-\sqrt{2\over 3})^3
{(\tanh{f}+1)^{{\sqrt{3\over 2}}(1-\sqrt{2\over 3})}\over
(\tanh{f}-1)^{{\sqrt{3\over 2}}(1+\sqrt{2\over 3})}},\ee
with $c_1$ an integration constant.
Now equations (\ref{eq1}) and (\ref{eq2}) can be integrated as follows:
\be \label{sol2}r+c_2={{\sqrt{3}}\over \z}\int
d\vphi {{e^{-\vphi}{\sqrt{1-\tanh^2{f(\vphi)}}}}}\ee
and
\be \label{sol3}A=-{1\over 6}{\sqrt{3\over 2}}\int d\vphi
\tanh{f(\vphi)}.\ee
The method therefore to find a solution 
is to invert (\ref{sol1}) for $\tanh{f}$
in terms of $\vphi$, substitute into (\ref{sol2}) 
and (\ref{sol3}) and integrate. Then,
to find the expression for $\vphi (r)$, invert (\ref{sol2}) and finally
use this to obtain the expression for $A(r)$.

The construction of the solution suited for an orbifold is described in
detail for example in \cite{sda}. Consistency of our flat domain wall
ansatz with the vanishing of the total 4 dimensional CC, amounts
to satisfying the following jump conditions:
\be 2\vphi '(r)=+{\partial T_{1}\over {\partial {\vphi}}}
(\vphi (r))\mid_{r=0}\label{jump1}\ee
\be 2A'(r)=-{1\over 6}T_{1}(\vphi (r))\mid_{r=0}\label{jump3}\ee
\be 2\vphi '(r)=-{\partial T_{2}\over {\partial {\vphi}}}(\vphi
(r))\mid_{r=\pi R}\label{jump2}\ee
\be 2A'(r)=+{1\over 6}T_{2}(\vphi (r))\mid_{r=\pi R}.\label{jump4}\ee
Equations (\ref{jump1}),
(\ref{jump2}), (\ref{jump3}) and (\ref{jump4}) is the system of
equations that has to be satisfied in a model with vanishing CC.
To satisfy these, we have the two integration constants $c_1$
and $c_2$, the size of the orbifold $R$ and
the discrete values of the $n_i$ that determine $\z^2$.
At the supersymmetric point,
we can rewrite (\ref{jump1})-(\ref{jump4}) in a more convenient form
using (\ref{elliptic}):
\be \cosh{f\bigl(\vphi(0)\bigr)}=\sqrt{3} \ee
\be \sinh{f\bigl(\vphi(0)\bigr)}=\sqrt{2} \ee
\be \cosh{f\bigl(\vphi(\pi R)\bigr)}=-\sqrt{3}\ee
\be \sinh{f\bigl(\vphi(\pi R)\bigr)}=-\sqrt{2}.\ee
A simple solution to the above can be found if 
we take $W=\z e^{\vphi}$, which is
a solution to (\ref{VW}). Then,
\be \cosh{f\bigl(\vphi(r)\bigr)}=\pm\sqrt{3}\;\;\;\;\; {\rm and}\;\;\;\;\;
\sinh{f\bigl(\vphi(r)\bigr)}=\pm\sqrt{2},\ee
and the equations of motion (\ref{sol2}) and (\ref{sol3}) 
can be solved easily, yielding
\be \vphi(r)=-\ln{(\z |r|+c)}\;\;\;\;\; {\rm and}\;\;\;\;\;
A(r)={1\over 6}{\ln{(\z |r|+c)}},\ee
with $c\equiv \z c_2$ the (only) integration constant.
This is a trivial solution, trivial in the sense that the jump
conditions are satisfied identically, without restriction on $\z$ and
the integration constant $c$.
In fact, since $\tanh{f\bigl(\vphi(r=0,\pi R)\bigr)}=\sqrt{2\over 3}$ at the
supersymmetric point, from (\ref{integ}) we see that at 
the supersymmetric point, this is actually the only possible solution.
\footnote{In general, even for compactifications on curved spaces such as 
type IIB on (squashed) $S^5$ for example, such a trivial solution is
always possible if the brane tension is taken to be $W$ \cite{kallosh}.}
Turning this around, we conclude that after supersymmetry breaking,
the simple ansatz $W=\z e^{\vphi}$ can not be used anymore to satisfy
(\ref{jump1})-(\ref{jump4}). But this is expected, since
in general it is not possible to satisfy the four jump equations with
the three parameters $\z$, $c$ and $R$ (even if $\z$ were continuous). Thus,
for a consistent model with zero cosmological constant away from the
supersymmetric point, we have to look for more general solutions to (\ref{VW}).

After supersymmetry breaking (on the brane), the brane tensions get renormalized
as \be T_1(\vphi)\rightarrow 2\z
e^{\vphi}(1+\e\psi_1(\z\e^{\vphi})) ,\;\;\;\;\;
T_2(\vphi)\rightarrow  2\z e^{\vphi}(1+\e\psi_2(\z\e^{\vphi}))\ee
with $\e $ being a small parameter characterizing the size of
supersymmetry breaking.  
The scalar potential on the other hand, at least at the 5D level, 
remains the same \cite{gaugino}.

The jump conditions therefore, after supersymmetry breaking become 
\be \label{jns1}\cosh{f\bigl(\vphi(0)\bigr)}=\sqrt{3}\Bigl(1+ \e
{d\over dx}(x\psi_1 (x))|_{x=x(0)} \Bigr) \ee
\be \label{jns2}\sinh{f\bigl(\vphi(0)\bigr)}=\sqrt{2} \Bigl(1+\e
\psi_1 (x)|_{x=x(0)} \Bigr)\ee \be
\label{jns3}\cosh{f\bigl(\vphi(\pi R)\bigr)}=-\sqrt{3}\Bigl(1+
\e
{d\over dx}(x\psi_2 (x))|_{x=x(\pi R)}\Bigr)\ee \be
\label{jns4}\sinh{f\bigl(\vphi(\pi R)\bigr)}=-\sqrt{2}\Bigl(1+\e
\psi_2 (x)|_{x=x(\pi R)}\Bigr),\ee
where $x=\z e^{\vphi}$.
Now let us recall that the difference of 
$\cosh f ~(\sinh f )$ from $\sqrt 3 ~(\sqrt 2)$
vanishes like $c_1$ which is of $O(\e )$. 
Thus we may write the above equations as
relations between $O(1)$ functions. Now in 
accordance with the general argument of
\cite{DeW}, the integration constants $c_1, c_2$ 
may be adjusted to satisfy say the first two equations above. 
In addition the distance $\pi R$ may be freely adjusted to satisfy 
one more matching condition leaving us with one 
more condition to satisfy. To satisfy
the fourth condition then requires a fine tuning 
of the parameters in the bulk potential.
In our case this means a fine tuning of the 
quantity $\zeta$. However, as noted earlier,
the latter is completely determined in terms of 
integers $n_i$ governing the fluxes
and the integers $d_{ijk}$ and $h_{11}$ of the 
Calabi-Yau manifold. There is no way these can be
chosen to satisfy this last equation since any 
change of these  integers results in a O(1) 
change of the functions $\psi_{1,2}$ and it would be a 
miracle if for any choice of the integers
the condition could be satisfied
\footnote{Of course one does not require
exact satisfaction of the matching conditions. 
Strictly speaking all we
need is that the cosmological constant on 
the brane be of $O(10^{-120}M_P^4)$. 
This would still imply an adjustment of parameters to 
this accuracy as explained in \cite{DeW} and is clearly
ruled out in the present situation.}. In other words there is 
no adjustable continuous parameter that could be tuned to satisfy the equation. 

Let us now discuss possible generalizations. 
One possible non-vanishing supersymmetric correction to the scalar
potential comes from the instanton gas obtained from wrapping M5-branes
around the Calabi-Yau manifold or M2-branes 
wrapping around 3-cycles\footnote{It is not
clear whether there are actually non-zero contributions 
due to transverse M2 branes parallel to the walls since 
there is no S-dual analog in the weak coupling 
i.e. the heterotic string limit.}. 
The potential would then be as in (\ref{VM5}) plus an analogous
contribution coming from M2-branes. 
Clearly, the original problem associated with the discreteness of the
fourth adjustable parameter still remains since we are still talking about
contributions that are completely determined by a set of integers.  

Another possible generalization would be to look for solutions to the 5D
equations of motion (\ref{eqX}) with $r$-dependent 
$b^i$. Such a solution, for the supersymmetric case, was found in
\cite{M}. After supersymmetry breaking, we would have to satisfy two
additional jump conditions for each $b^i$, which is possible, since we
get two additional integration constants for each $b^i$ from their
equations of motion. However, as before, we still have one fine tuning to do
and the previous argument, that since there is no
continuous parameter there is no possibility of doing this, still applies.

%%%%%%%%%%%%%%%%%%%%%%%%%%%%%%%%%%%%%%%%%%%%%%%%%%%%%%%%%%%%%%%%%%%%%%%%%%%%
\section{Conclusions}
%%%%%%%%%%%%%%%%%%%%%%%%%%%%%%%%%%%%%%%%%%%%%%%%%%%%%%%%%%%%%%%%%%%%%%%%%%%%

We have discussed in this paper the two brane scenario coming from
the Horava-Witten theory compactified on a Calabi-Yau manifold
obtained by Lukas et al \cite{M}. Those authors discussed the 
supersymmetry preserving case when there is sufficient degeneracy
in the system of equations to satisfy the matching conditions without
any adjustment of integration constants. When supersymmetry is
broken however, as has been discussed in earlier work, we need four
adjustable parameters (integration constants).
We have shown explicitly that how these  four parameters 
arise in the case at hand but that since one of the 
parameters necessarily takes discrete values, it does 
not seem possible to tune the cosmological constant to zero (or a small value).

The situation discussed in this paper is to be contrasted with the
one obtained for  compactifications on Ricci non-flat manifolds,
as for example in IIB string theory on a (squashed) $S^5$, where
one has two exponentials in the potential for the breathing mode.
In these compactifications, besides a discrete parameter like $\z$
which is also present, (coming from turning on five-form flux)
there is an additional continuous parameter not present in
Calabi-Yau compactifications. It is associated with the Ricci
curvature of the compact space. Thus, in such cases, it is in
principle always possible to satisfy the jump conditions and
readjust integration constants by an arbitrarily small quantity
after supersymmetry breaking to get flat brane solutions i.e. a
cosmological constant that is zero\footnote{Or indeed of order
$O((10^{-3} eV)^4$ as seems to be fashionable these days!} after
supersymmetry breaking. Of course in the absence of a theory of
integration constants one cannot claim to have solved the
cosmological constant problem even in the case of
compactifications on Ricci-non-flat manifolds. All one could claim
there is that one has sufficient freedom to get a flat space
solution. In the Ricci flat case what we have argued is that this
freedom does not seem to exist. 

%%%%%%%%%%%%%%%%%%%%%%%%%%%%%%%%%%%%%%%%%%%%%%%%%%%%%%%%%%%%%%%%%

%%%%%%%%%%%%%%%%%%%%%%%%%%%%%%%%%%%%%%%%%%%%%%%%%%%%%%%%%%%%%%%%%
\vskip 1cm
{\bf Acknowledgements}
\vskip .5cm
%%%%%%%%%%%%%%%%%%%%%%%%%%%%%%%%%%%%%%%%%%%%%%%%%%%%%%%%%%%%%%%%%%%%%%%%%%%

This work is partially supported by the Department of
Energy contract No. DE-FG02-91-ER-40672.

%%%%%%%%%%%%%%%%%%%%%%%%%%%%%%%%%%%%%%%%%

%%%%%%%%%%%%%%%%%%%%%%%%%%%%%%%%%%%%%%%%%%%%%%%%%%%%%%%%%%%%%%%%%%%%%%%%%%%%
\end{document}